\documentclass[prl,aps,showpacs,twocolumn,twoside,superscriptaddress]{revtex4}

\usepackage{amsfonts}
\usepackage{amsmath}
\usepackage{amssymb}
\usepackage{graphicx}
\usepackage{mathrsfs}

\begin{document}
\title{Distillation of Multipartite Entanglement by Complementary Stabilizer Measurements}

\author{Akimasa Miyake}
\email{akimasa.miyake@uibk.ac.at}
\affiliation{\mbox{Institute for Theoretical Physics, University of Innsbruck, 
Technikerstra{\ss}e 25, A-6020 Innsbruck, Austria}}
\affiliation{\mbox{Institute for Quantum Optics and Quantum Information, 
Austrian Academy of Sciences, Innsbruck, Austria}}
\affiliation{\mbox{Graduate School of Science, University of Tokyo, Hongo 7-3-1 
Bunkyo-ku, Tokyo 113-0033, Japan}}

\author{Hans J. Briegel}
\email{hans.briegel@uibk.ac.at}
\affiliation{\mbox{Institute for Theoretical Physics, University of Innsbruck, 
Technikerstra{\ss}e 25, A-6020 Innsbruck, Austria}}
\affiliation{\mbox{Institute for Quantum Optics and Quantum Information, 
Austrian Academy of Sciences, Innsbruck, Austria}}

\date{June 11, 2005} 

%
%
\begin{abstract}
We propose a scheme of multipartite entanglement distillation driven 
by a complementary pair of stabilizer measurements, to distill directly 
a wider range of states beyond the stabilizer code states 
(such as the Greenberger-Horne-Zeilinger states).
We make our idea explicit by constructing a recurrence protocol 
for the 3-qubit W state 
$\frac{1}{\sqrt{3}}(|001\rangle +|010\rangle +|100\rangle)$.
Noisy W states resulting from typical decoherence can be directly
purified in a few steps, if their initial fidelity is larger than a threshold.
For general input mixed states, we observe distillations to 
hierarchical fixed points, i.e., not only to the W state but also to
the 2-qubit Bell pair, depending on their initial entanglement. 

\end{abstract}

\maketitle

%
%
{\it Introduction.--}
We recognize, through a decade of research, that entanglement is
indispensable to execute quantum information processing (QIP),
such as quantum computation and multi-party quantum communication.
A persistent challenge is to maintain multipartite entanglement against 
decoherence. 
In this Letter, we enlarge the present applicability of a key 
technique, entanglement distillation \cite{bennett96-1,bennett96-2}, 
to genuine multipartite entangled states called the W states
\cite{coffman00}.

The W state, $\tfrac{1}{\sqrt{n}}(|0\ldots 01\rangle + |0\ldots 10\rangle + 
\cdots + |10\ldots 0\rangle)$, i.e., the equal superposition of 
``single-excitation'' basis vectors in $n$ qubits is a tolerant {\em resource} 
against decoherence and loss of qubits.
It is quite robust \cite{briegel01,dur01},
because it can be compared to a symmetric web consisting of only
pairwise entanglement \cite{coffman00,koashi00}.
This state is also a Dicke state of the total spin operators 
$\vec{J}^2$ and $J_z$ with eigenvalues $\tfrac{n}{2}(\tfrac{n}{2}+1)$ and 
$\tfrac{n}{2}-1$ respectively, due to the permutation symmetry.
Thus the W state is often available more easily than the 
Greenberger-Horne-Zeilinger (GHZ) state.
The 3-qubit W state has been created in optical systems \cite{eibl04} and 
ion traps \cite{roos04}, and can be prepared according to several proposals 
in coupled quantum dots, critical spin chains, etc.
Furthermore, the W state can be said to be essentially different from
most multipartite entangled states known in the applications to QIP, 
in the following sense.

Basic {\em software techniques} to circumvent decoherence have been proposed,
and implemented experimentally for prototypes. 
For example, entanglement distillation (or, purification) 
\cite{bennett96-1,bennett96-2} is a tool to extract high-fidelity entangled
states from a larger ensemble of noisy ones.
Quantum error correction codes \cite{qecc,gottesman97}
are a way to protect entanglement from small numbers of errors.
The latter can be formulated in terms of the stabilizer, i.e., as simultaneous 
eigenspaces of commuting ``multilocal'' Pauli operators.
Note that, if all stabilized eigenspaces are 1-dimensional state vectors, 
they are often called stabilizer states \cite{gottesman98} or graph states 
(up to local unitaries) \cite{briegel01,schlingemann02}.
The Calderbank-Shor-Steane (CSS) code \cite{qecc} is 
defined by the stabilizer group which consists of only two kinds of generators:
multilocal bit-flip $X$ operators and phase-flip $Z$ operators.
In fact, beyond the ``bipartite'' distillation protocols 
\cite{bennett96-1,bennett96-2,deutsch96} for the Bell pairs,
direct distillation of multipartite entanglement is so far possible
just for the CSS stabilizer (or, two-colorable graph) states by 
the protocol in Refs.~\cite{dur03,chen04}, which extended earlier results for 
GHZ states \cite{murao98,maneva00}.
Since the W state is not a stabilizer state, there has been no protocol
to distill it directly.

{\it Main idea.--}
We propose an entanglement distillation protocol 
that extracts directly a multipartite non-stabilizer (non-graph) state, 
specifically the 3-qubit W state.
Our idea is to apply {\em local} measurements of the stabilizer (whose
nonlocal counterpart acting at different parties stabilizes the target state),
assuming that the target state belongs to a basis of equivalent 
entangled states.
Note that such a basis, similar to the Bell basis, exists for a wider
range of multipartite states than stabilizer states.
We need $n$ copies of the input state for the $n$-qubit case, to apply 
stabilizer measurements locally. 
In this manner, we can improve the fidelity, and attain the target state 
as a fixed point of the protocol.
Note that if the target state is not the stabilizer state, {\em local} 
depolarization or twirling (over the single copy) which keeps the target state 
invariant seems impossible in general.
Thus, we do not make the mixed states diagonal, i.e., a classical mixture of 
the basis states.
It implies we cannot reduce the task to a ``classical problem'' that consists
in extracting entropy from the binary strings of the stabilizer eigenvalues,
as is possible by bilateral {\sc cnot} operations in all the known protocols.
Nevertheless, by virtue of complementary stabilizer measurements which exchange 
the amplified components, our protocol works without local depolarization.
The feature is favorable in efficiency, and analogous to 
the Oxford protocol \cite{deutsch96}.

Direct distillation of multipartite entanglement has several potential merits.
In the case of CSS states such as the GHZ states, multipartite distillation 
was shown to be more efficient than the bipartite strategy which consists
of distillation of Bell pairs and their connection \cite{murao98,dur03}.
Under imperfect operations, the achievable fidelity can be higher 
\cite{dur03,kruszynska05}.
Also, the threshold for distillability may be tighter than that by
indirect methods.

{\it W basis and its stabilizer group.--}
To make our idea explicit, we construct a recurrence protocol
for the 3-qubit W state $|W^{000}\rangle = \tfrac{1}{\sqrt{3}}(|001\rangle 
+ |010\rangle + |100\rangle)^{ABC}$, distributed over Alice, Bob, and Carol.
We denote the Pauli matrices, operating on the $j$-th qubit at the party 
$l$, as $X_{j}^{l}$, $Y_{j}^{l}$, and $Z_{j}^{l}$
along with the identity $\openone_{j}^{l}$. 
To distinguish the non-local tensor structure of the multiple Hilbert spaces
controlled by different parties and the local tensor structure 
at a single party, we use the superscripts $l$ as the non-local indices, 
and the subscripts $j$ as the local ones.
We define the Hadamard operation by $H = \tfrac{1}{\sqrt{2}}(X + Z)$, 
the 2-qubit swap operation by $\mbox{\sc swap}: |k k'\rangle \mapsto
|k' k\rangle$ $(k,k' = 0,1)$ in the computational basis, and
a 3-qubit unitary operation $V$, which leaves $|000\rangle$ and 
$|111\rangle$ unchanged, but exchanges the others in such a way that 
$|001\rangle \leftrightarrow |110\rangle$, 
$|010\rangle \leftrightarrow |101\rangle$, and
$|100\rangle \leftrightarrow |011\rangle$.

Let us introduce a complete orthonormal basis, called the W basis here,
where each basis state $|W^{k_1 k_2 k_3}\rangle$ $(k_1,k_2,k_3 =0,1)$ 
has entanglement equivalent to the W state $|W^{000}\rangle$.
This is because basis states transform into each other by the local 
unitary operations in Table~\ref{tab:wbasis}.
The W basis can be obtained from the computational basis
acted on by the 3-qubit unitary operation
$U^{\rm W basis} = \tfrac{1}{\sqrt{3}}(\openone^{A}Z^{B}X^{C} + 
Z^{A}X^{B}\openone^{C} + X^{A}\openone^{B}Z^{C})$, i.e.,
$|W^{k_1 k_2 k_3}\rangle = U^{\rm W basis}|k_1 k_2 k_3\rangle$.
It is convenient to identify the stabilizer for the W basis.
To satisfy $K_{j}|W^{k_1 k_2 k_3}\rangle = (-1)^{k_j}|W^{k_1 k_2 k_3}\rangle$,
three generators $K_j$ are determined as
\begin{align}
\begin{split}
K_1^{(ABC)} \!\!\!\! &= \tfrac{1}{3}(2 X^{A}X^{B}Z^{C} + 2 Y^{A}Z^{B}Y^{C} + 
Z^{A}\openone^{B}\openone^{C}) , \\
K_2^{(ABC)} \!\!\!\! &= \tfrac{1}{3}(2 Z^{A}X^{B}X^{C} + 2 Y^{A}Y^{B}Z^{C} + 
\openone^{A}Z^{B}\openone^{C}) , \\
K_3^{(ABC)} \!\!\!\! &= \tfrac{1}{3}(2 X^{A}Z^{B}X^{C} + 2 Z^{A}Y^{B}Y^{C} + 
\openone^{A}\openone^{B}Z^{C}) .
\end{split}
\end{align}
We emphasize, by the superscript $(ABC)$, that the stabilizers
are not local. Note that later we measure {\em locally} 
the stabilizers, which will be denoted, e.g. for Alice, as $K_1^{(A)} = 
\tfrac{1}{3}(2 X_{1}^{A}X_{2}^{A}Z_{3}^{A} + 2 Y_{1}^{A}Z_{2}^{A}Y_{3}^{A} + 
Z_{1}^{A}\openone_{2}^{A}\openone_{3}^{A})$, and
all 3 qubits specified by the {\em subscripts} belong to Alice.
The stabilizer group consists of eight commuting elements,
$\{\openone, K_1, K_2, K_3, K_1 K_2, K_1 K_3, K_2 K_3, K_1 K_2 K_3 \}$,
where
\begin{align}
\begin{split}
K_1 K_2 &= \tfrac{1}{3}(2 \openone_{1}X_{2}X_{3} + 2 Y_{1}\openone_{2}Y_{3} - 
Z_{1} Z_{2} \openone_{3}) , \\
K_1 K_3 &= \tfrac{1}{3}(2 X_{1}X_{2}\openone_{3} + 2 \openone_{1}Y_{2}Y_{3} -
Z_{1}\openone_{2}Z_{3}) , \\
K_2 K_3 &= \tfrac{1}{3}(2 X_{1}\openone_{2}X_{3} + 2 Y_{1}Y_{2}\openone_{3} -
\openone_{1}Z_{2}Z_{3}) , \\
K_1 K_2 K_3 &= -Z_{1}Z_{2}Z_{3}. \\
\end{split}
\end{align}

\begin{table}[t]
\begin{tabular}{c|c|c}
$k_1 k_2 k_3$ & $|W^{k_1 k_2 k_3}\rangle$ & 
$|W^{000}\rangle \mapsto |W^{k_1 k_2 k_3}\rangle$ \\
\hline 
$000$ & $\tfrac{1}{\sqrt{3}}(|001\rangle + |010\rangle + 
|100\rangle)$& $\openone^{A}\openone^{B}\openone^{C}$ \\
$001$ & $\tfrac{1}{\sqrt{3}}(|000\rangle + |011\rangle -
|101\rangle)$ & $Z^{A}\openone^{B}X^{C}$ \\
$010$ & $\tfrac{1}{\sqrt{3}}(-|011\rangle + |000\rangle +
|110\rangle)$ & $\openone^{A}X^{B}Z^{C}$ \\
$011$ & $\tfrac{1}{\sqrt{3}}(-|010\rangle + |001\rangle -
|111\rangle)$ & $Z^{A}X^{B}(-iY^{C})$ \\
$100$ & $\tfrac{1}{\sqrt{3}}(|101\rangle - |110\rangle + 
|000\rangle)$ & $X^{A}Z^{B}\openone^{C}$ \\
$101$ & $\tfrac{1}{\sqrt{3}}(|100\rangle - |111\rangle -
|001\rangle)$ & $(-iY^{A})Z^{B}X^{C}$ \\
$110$ & $\tfrac{1}{\sqrt{3}}(-|111\rangle - |100\rangle + 
|010\rangle)$ & $X^{A}(-iY^{B})Z^{C}$ \\
$111$ & $\tfrac{1}{\sqrt{3}}(-|110\rangle - |101\rangle -
|011\rangle)$ & $(-iY^{A})(-iY^{B})(-iY^{C})$  
\end{tabular}
\caption{The 3-qubit W basis. Local unitary operations that map 
$|W^{000}\rangle$ to $|W^{k_1 k_2 k_3}\rangle$ are shown in the third column.}
\label{tab:wbasis}
\end{table}

{\it W state distillation protocol.--}
Our protocol consists of two subprotocols: 
${\mathscr P}$ and its dual $\bar{\mathscr P}$. In both,
three input copies are mapped into one output copy to define 
a simple recurrence.
Generally, any mapping to a smaller subsystem can be considered.
We assume, without loss of generality, that 
the $|W^{000}\rangle\langle W^{000}|$ component of the input mixed states 
$\rho$ is the largest among the diagonal elements in the W basis
(otherwise we can relabel the computational basis by a local unitary
operation in Table~\ref{tab:wbasis}).
We define the fidelity $F$ of $\rho$ by
$F = \langle W^{000}|\rho |W^{000}\rangle$.

{\it Protocol ${\mathscr P}$:}
1) Every party ($l$ = A, B, or C) applies the local measurement of two 
stabilizers $K_1^{(l)} K_2^{(l)}$ and $K_1^{(l)} K_3^{(l)}$ over  
the input state $\gamma$ (= $\rho_{\rm in}^{\otimes 3}$) of three copies, 
and obtains the 2-bit outcomes  
${\mathbf m}^{(l)} = [m_1^{(l)},m_2^{(l)}]$. 
2) Informing their outcomes by two-way classical communication, 
parties select coincident outcomes 
${\mathbf m}^{(A)} = {\mathbf m}^{(B)} = {\mathbf m}^{(C)} = [0,1] \;
(\equiv {\bf 1})$, $[1,0] \;(\equiv {\bf 2})$, or $[1,1] \;(\equiv {\bf 3})$.
Otherwise they discard three copies.
3) For the coincident outcomes, each party transforms {\em locally}  
her/his state into a 1-qubit subsystem by the following ``majority rule''.
If ${\mathbf m}^{(l)} = {\bf 1}$, $P_{\bf 1}^{l}: 
|W_{001}\rangle^{l} \mapsto |0\rangle^{l}$, 
$|W_{110}\rangle^{l} \mapsto |1\rangle^{l}$;
if ${\mathbf m}^{(l)} = {\bf 2}$, $P_{\bf 2}^{l}: 
|W_{010}\rangle^{l} \mapsto |0\rangle^{l}$, 
$|W_{101}\rangle^{l} \mapsto |1\rangle^{l}$; and
if ${\mathbf m}^{(l)} = {\bf 3}$, $P_{\bf 3}^{l}: 
|W_{100}\rangle^{l} \mapsto |0\rangle^{l}$, 
$|W_{011}\rangle^{l} \mapsto |1\rangle^{l}$.

Mathematically, the stabilizer measurement 
$M_{{\mathbf m}^{(l)}}^{(l)}$ of the party $l$ is written by,
\begin{equation}
\label{eq:m}
M_{{\mathbf m}^{(l)}}^{(l)} \!\!=\!\! 
\tfrac{1}{4} 
\left(\openone + (-1)^{m_1^{(l)}} K_1^{(l)}K_2^{(l)}\right)\!\!\!
\left(\openone + (-1)^{m_2^{(l)}} K_1^{(l)}K_3^{(l)}\right),
\end{equation}
with the completeness condition 
$\sum_{{\mathbf m}^{(l)}} M_{{\mathbf m}^{(l)}}^{(l)\dag}
M_{{\mathbf m}^{(l)}}^{(l)} = \openone$.
Note that $M_{{\mathbf m}^{(l)}}^{(l)}$ acts on 3 qubits of the party
$l$, and it is a projector to the {\em local} W basis vectors, for example 
if ${\mathbf m}^{(l)}={\bf 1}$, $M_{\bf 1}^{(l)}=
|W_{001}\rangle^{l}\langle W_{001}| + |W_{110}\rangle^{l}\langle W_{110}|$.
By the selection of desired coincident outcomes ${\bf m}$, $\mathscr P$ maps 
the input state $\gamma$ (= $\rho_{\rm in}^{\otimes 3}$) to 
the one-copy state $\rho'$ given by
\begin{equation}
\label{eq:p} 
\rho' = \!\!\!\!\!
\sum_{{\mathbf m}={\bf 1,2,3}} \!\!\!\!\!
PM_{\mathbf m}^{(A)}PM_{\mathbf m}^{(B)}PM_{\mathbf m}^{(C)} \gamma
PM_{\mathbf m}^{(A)\dag}PM_{\mathbf m}^{(B)\dag}
PM_{\mathbf m}^{(C)\dag},
\end{equation}
with the success probability ${\rm tr}(\rho')$.
We normalize the state as $\rho_{\rm out}=\rho'/{\rm tr} (\rho')$ 
for the next recurrence step.

\begin{figure}[t]
\begin{minipage}{4.5cm}
\begin{center}
\includegraphics[width=4.7cm,clip]{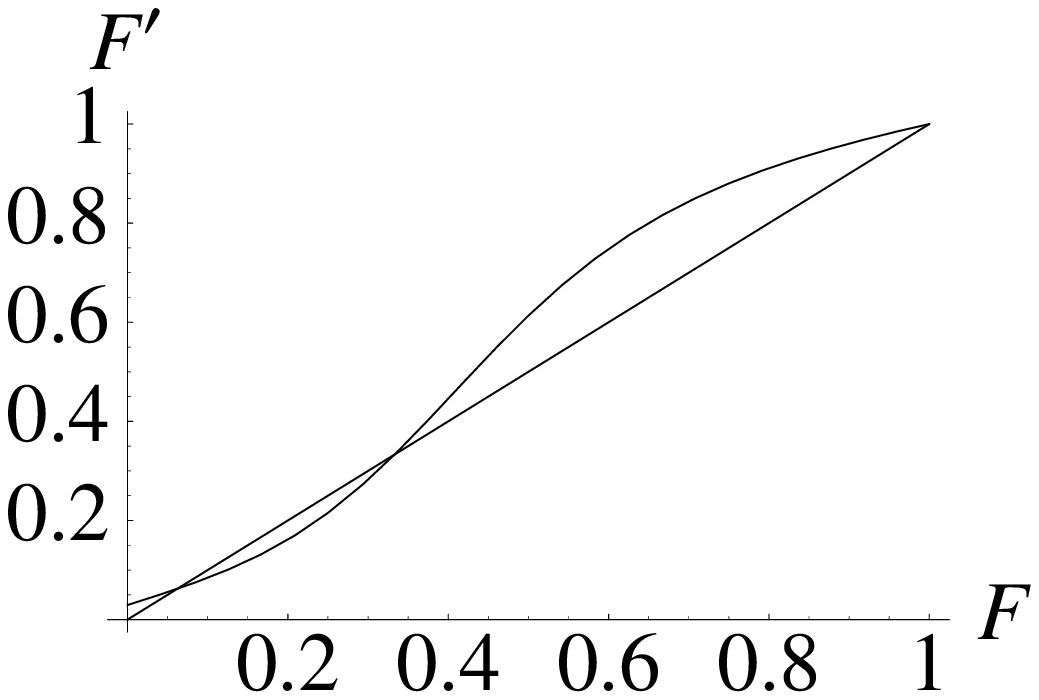}
\end{center}
\end{minipage}
\begin{minipage}{4.0cm}
\begin{center}
\includegraphics[width=4.0cm,clip]{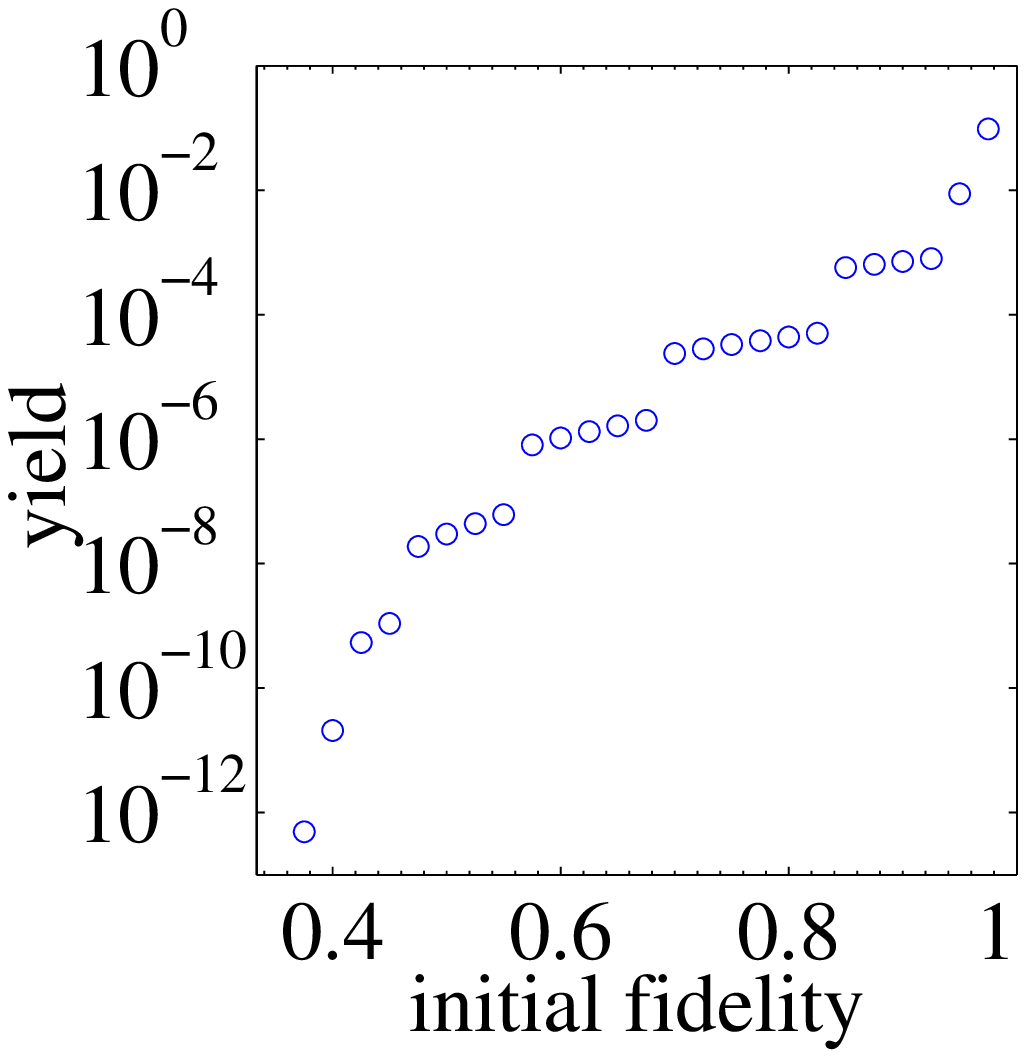}
\end{center}
\end{minipage}
\caption{The distillation curve (left) of ${\mathscr P}$ for locally 
dephased W states, and the yield (right) after ${\mathscr P}$ achieves 
$F \geq 0.99$. Note that the region of $F \in [0,\tfrac{1}{3})$ is not
physical. }
\label{fig:lodephase}
\end{figure}

%
Before describing the whole protocol including $\bar{\mathscr P}$,
we illustrate analytically how ${\mathscr P}$ works. 
Suppose the perfect W state is distributed to three parties, 
but suffers typical decoherence as described by 
the local dephasing channel 
${\mathcal D}^{l}(\rho) = \tfrac{1}{2}((1+\mu)\rho + (1-\mu)Z^{l}\rho Z^{l} )$ 
with the same channel reliability $\mu \in [0,1]$.
Three parties initially share a noisy W state 
$\sigma(F) = {\mathcal D}^{A}{\mathcal D}^{B}{\mathcal D}^{C}
(|W^{000}\rangle\langle W^{000}|)$, which is not diagonal in the W basis, 
but is parametrized uniquely by  
$F = \tfrac{1}{3}(1+ 2 \mu^2) \in [\tfrac{1}{3},1]$. 
A straightforward calculation shows that  ${\mathscr P}$ maps three copies 
$\sigma(F)^{\otimes 3}$ to one copy $\sigma(F')$ with the higher fidelity 
$F'$ such that
\begin{equation}
\label{eq:fid_lodephase}
F' = \frac{\tfrac{25}{81}F^3 + \tfrac{1}{18}F(1-F)^2 + 
\tfrac{1}{324}(1-F)^3}{\tfrac{25}{81}F^3 + \tfrac{1}{9}F^2 (1-F)
+ \tfrac{2}{27}F(1-F)^2 + \tfrac{17}{162}(1-F)^3}.
\end{equation}
Eq.~(\ref{eq:fid_lodephase}) suggests a recurrence  
seen in the distillation curve of Fig.~\ref{fig:lodephase}.
Since $F$ lies in $[\tfrac{1}{3},1]$, we prove analytically that $F=1$, 
corresponding to the W state, is the {\em attractive} fixed point and 
$F=\tfrac{1}{3}$ is the repulsive one. 
We find that for any locally dephased W state except 
$F=\tfrac{1}{3}$, ${\mathscr P}$ restores it with a few steps.
Indeed, this threshold coincides with a {\em necessary} condition for 
distillability by the partial transposition criterion \cite{peres96,dur99}.
Since the mixed state ${\mathcal T^{l}} (\sigma(F))$, partially transposed 
for any party $l$ (i.e., bipartition), has a negative 
eigenvalue only for $F > \tfrac{1}{3}$, there is no chance
to distill entanglement in $F=\tfrac{1}{3}$.
In Fig.~\ref{fig:lodephase}, the yield (i.e., the ratio of the number of 
surviving copies to that of used copies) of ${\mathscr P}$ after $F$ reaches 
at least 0.99 is also shown.
The ``stairs'' of the yield come from the difference in the number of 
recurrence steps.

For more general noises, we need $\bar{\mathscr P}$ which has 
the similar structure as ${\mathscr P}$ but
employs complementary observables $\bar{K}^{(l)}_{j} = 
\Lambda^{l \dag} K^{(l)}_{j} \Lambda^{l}$, where
$\Lambda^l = H_1^l H_2^l H_3^l \mbox{\sc swap}_{13}^{l}$.
Two measurement bases $|W_{k_1 k_2 k_3}\rangle$ in ${\mathscr P}$ and 
$|\bar{W}_{k'_1 k'_2 k'_3}\rangle = \Lambda^{\dag}|W_{k'_1 k'_2 k'_3}\rangle$ 
in $\bar{\mathscr P}$ are complementary (also called {\em mutually unbiased} 
\cite{wootters89}), i.e., 
$|\langle W_{k_1 k_2 k_3}|\bar{W}_{k'_1 k'_2 k'_3}\rangle|^2 = \tfrac{1}{8}$.

{\it Dual Protocol $\bar{\mathscr P}$:}
0) Every party ($l$ = A, B, or C) applies 
$V^l$ to change the {\em local} computational basis.
The input $\gamma$ is modified to $\bar{\gamma} = V^A V^B V^C 
\rho_{\rm in}^{\otimes 3} V^{A \dag} V^{B \dag} V^{C \dag}$.
1) She/he applies the local measurement of two dual stabilizers 
$\bar{K}_1^{(l)}\bar{K}_2^{(l)}$ and $\bar{K}_1^{(l)}\bar{K}_3^{(l)}$
on $\bar{\gamma}$, and obtains the 2-bit outcomes $\bar{\mathbf m}^{(l)}$.
2) By two-way classical communication, parties select the coincident
outcome $\bar{\mathbf m}^{(A)}=\bar{\mathbf m}^{(B)} = \bar{\mathbf m}^{(C)} 
= [0,0] \;(\equiv {\bf 0})$.
3) Each party transforms locally the state into a 1-qubit subsystem in 
the manner opposite to ${\mathscr P}$; $\bar{P}_{\bf 0}^l : 
|\bar{W}_{000}\rangle^l \mapsto H |1\rangle^l$,
$|\bar{W}_{111}\rangle^l \mapsto H |0\rangle^l$.

In brief, in $\bar{\mathscr P}$, we replace all operators in Eqs.~(\ref{eq:m}) 
and (\ref{eq:p}) by their ``barred'' dual operators.
The complete distillation procedure for general mixed states consists of
the sequential application of either ${\mathscr P}$ or $\bar{\mathscr P}$ 
where, in every recurrence step, we select one of the subprotocols which gives 
the higher fidelity in the output.
There seems to be no simple formula for the sequence, but
we can determine the sequence if we know the initial 
input state $\rho$ before distillation, e.g., by state tomography.
Also note that although this combination of ${\mathscr P}$ and
$\bar{\mathscr P}$ can reach numerically the region where $F > 0.999$ , 
precisely speaking $F=1$ is the fixed point of 
${\mathscr P}$ but is not that of $\bar{\mathscr P}$.

\begin{figure}[t]
\begin{minipage}{3.9cm}
\begin{center}
\includegraphics[width=3.9cm,clip]{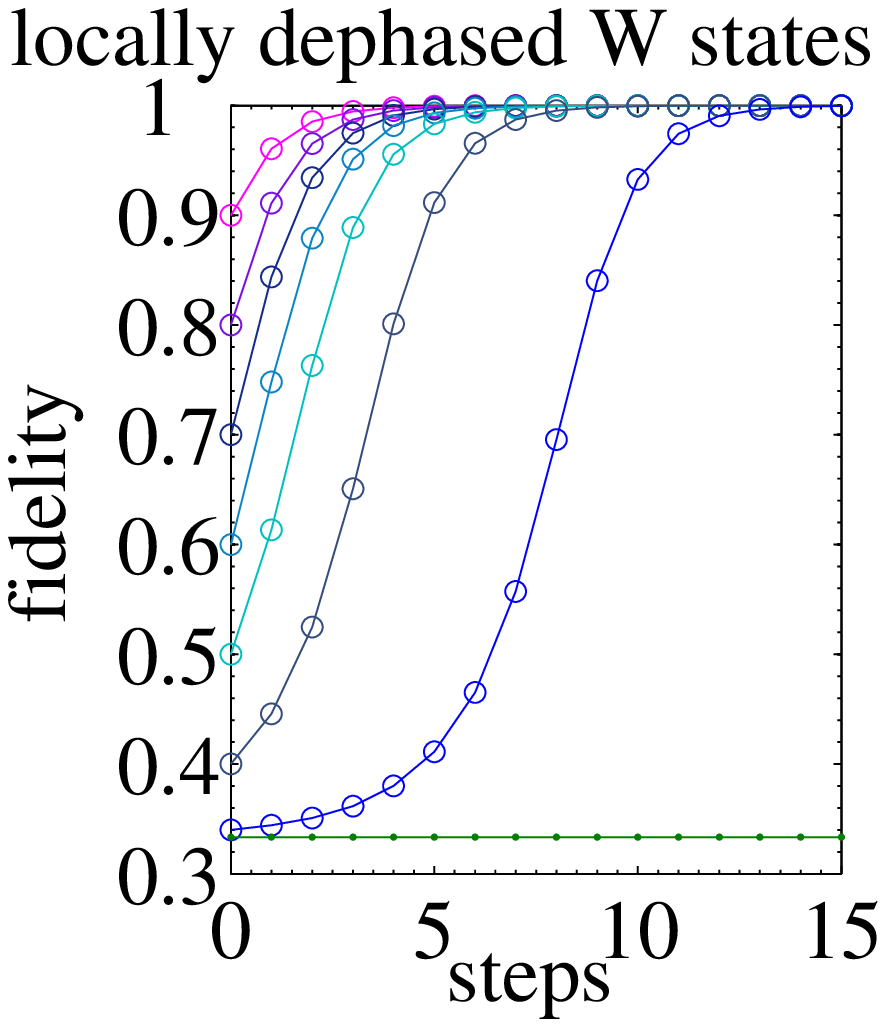}
\end{center}
\end{minipage}
\begin{minipage}{4.6cm}
\begin{center}
\includegraphics[width=4.6cm,clip]{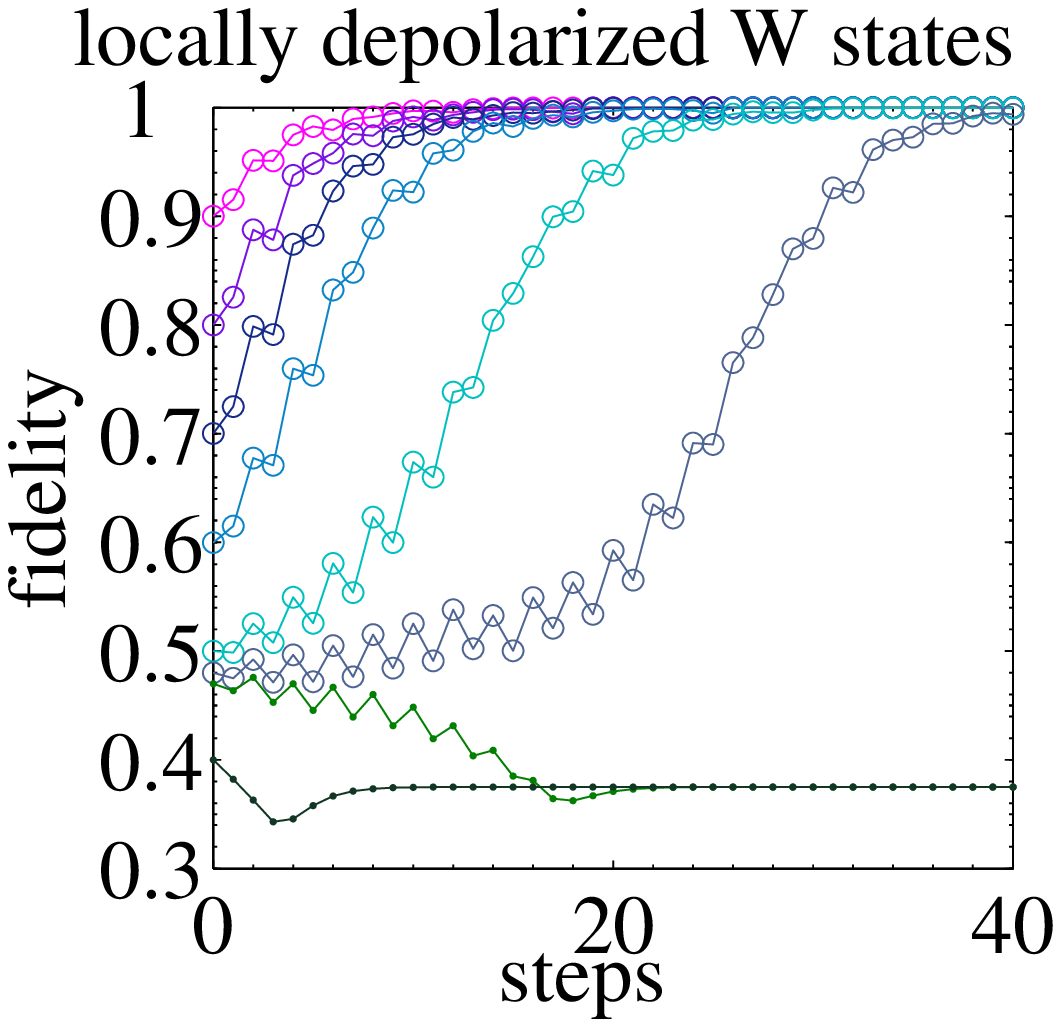}
\end{center}
\end{minipage}
\caption{Noisy W states subjected by the local dephasing (left) or 
local depolarizing (right) channel can be retrieved by ${\mathscr P}$ and 
$\bar{\mathscr P}$, if $F$ is initially larger than $\tfrac{1}{3}$ or $0.48$,
respectively. Note that this is actually accomplished by ${\mathscr P}$ alone
for the local dephasing case.}
\label{fig:fid_lonoise}
\end{figure}

%
%
Hereafter, we show that, under the sequential application of ${\mathscr P}$ and 
$\bar{\mathscr P}$, the W state can be distilled from arbitrary
mixed states if, roughly speaking, $F$ is sufficiently large.
First, consider another typical decoherence 
such as the local depolarizing channel (white noise) 
${\mathcal E}^{l}(\rho) = \mu\rho + 
\tfrac{1-\mu}{4}(\rho + X^{l}\rho X^{l} + Y^{l}\rho Y^{l} + Z^{l}\rho Z^{l})$,
and the input state
$\rho_{\rm in}(F)={\mathcal E}^{A}{\mathcal E}^{B}{\mathcal E}^{C}
(|W^{000}\rangle\langle W^{000}|)$ with  
$F=\tfrac{1}{24}(3+\mu + 9\mu^2 + 11\mu^3) \in [\tfrac{1}{8},1]$.
Although the locally depolarized W state does not remain in the same form 
under our protocol, we can still determine a threshold for distillability.
As seen in Fig.~\ref{fig:fid_lonoise}, if initially $F \gtrsim 0.48$, 
we distill the W state, and otherwise we  
have an undistillable mixed state $\chi = 
\frac{1}{2}(|\varphi\rangle\langle\varphi | + 
|\varphi' \rangle\langle\varphi' |)$, where 
$|\varphi \rangle = \tfrac{1}{2}
(|001\rangle + |010\rangle + |100\rangle - |111\rangle)$ and
$|\varphi'\rangle = \tfrac{1}{2}
(-|000\rangle + |011\rangle + |101\rangle + |110\rangle)$, as 
another fixed point with $F=\tfrac{3}{8}$.
This threshold is stricter than the necessary condition $F \gtrsim 0.36$ by 
the partial transpose criterion \cite{peres96,dur99}.
Note that the progress of the protocol is not described by a single parameter,
and $F$ is not monotonic any more.
A nonmonotonic behavior of $F$ was also seen in the bipartite 
distillation without depolarization \cite{deutsch96}.
However, as a long-term behavior, $F$ is increasing for the distillable
cases and can be used for visualization of the progress.

%
%
\begin{figure}[t]
\begin{minipage}{3.7cm}
\begin{center}
\includegraphics[width=3.7cm,clip]{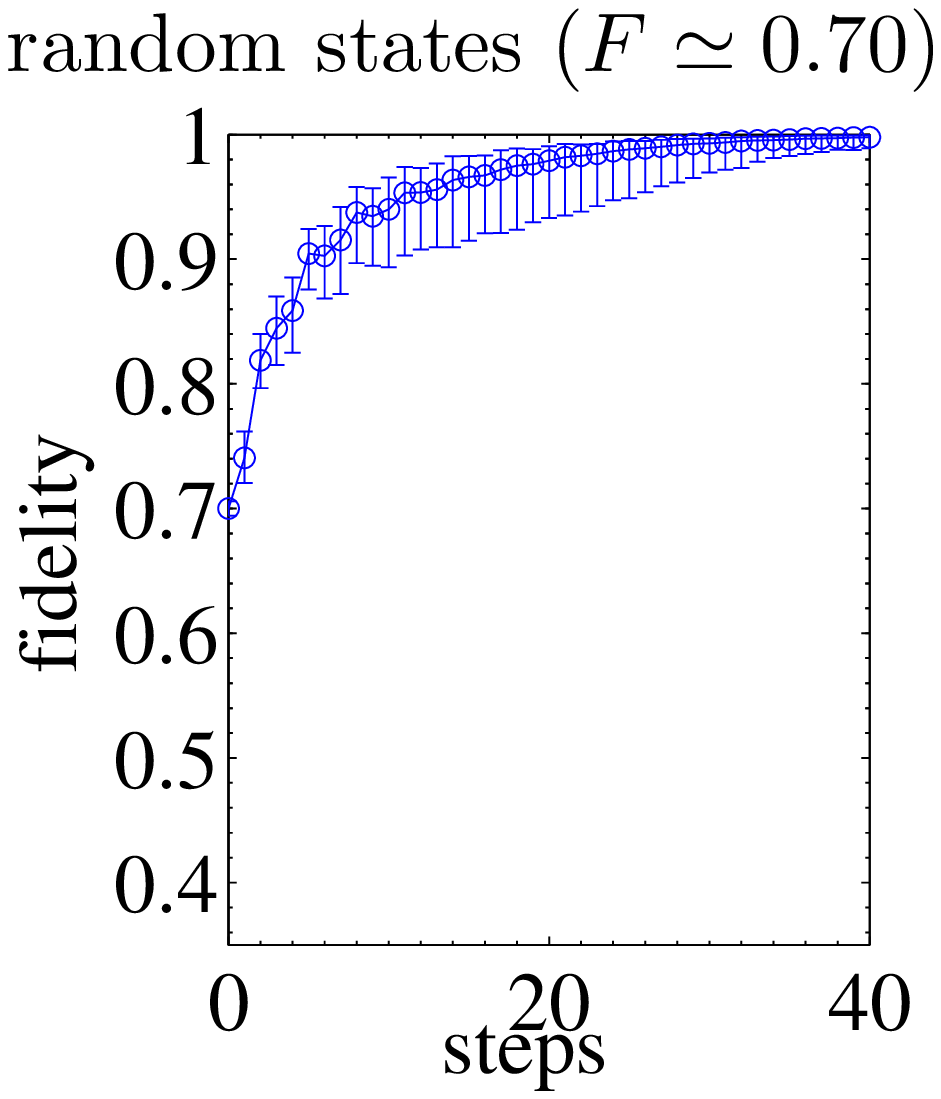}
\end{center}
\end{minipage}
\begin{minipage}{4.8cm}
\begin{center}
\includegraphics[width=4.8cm,clip]{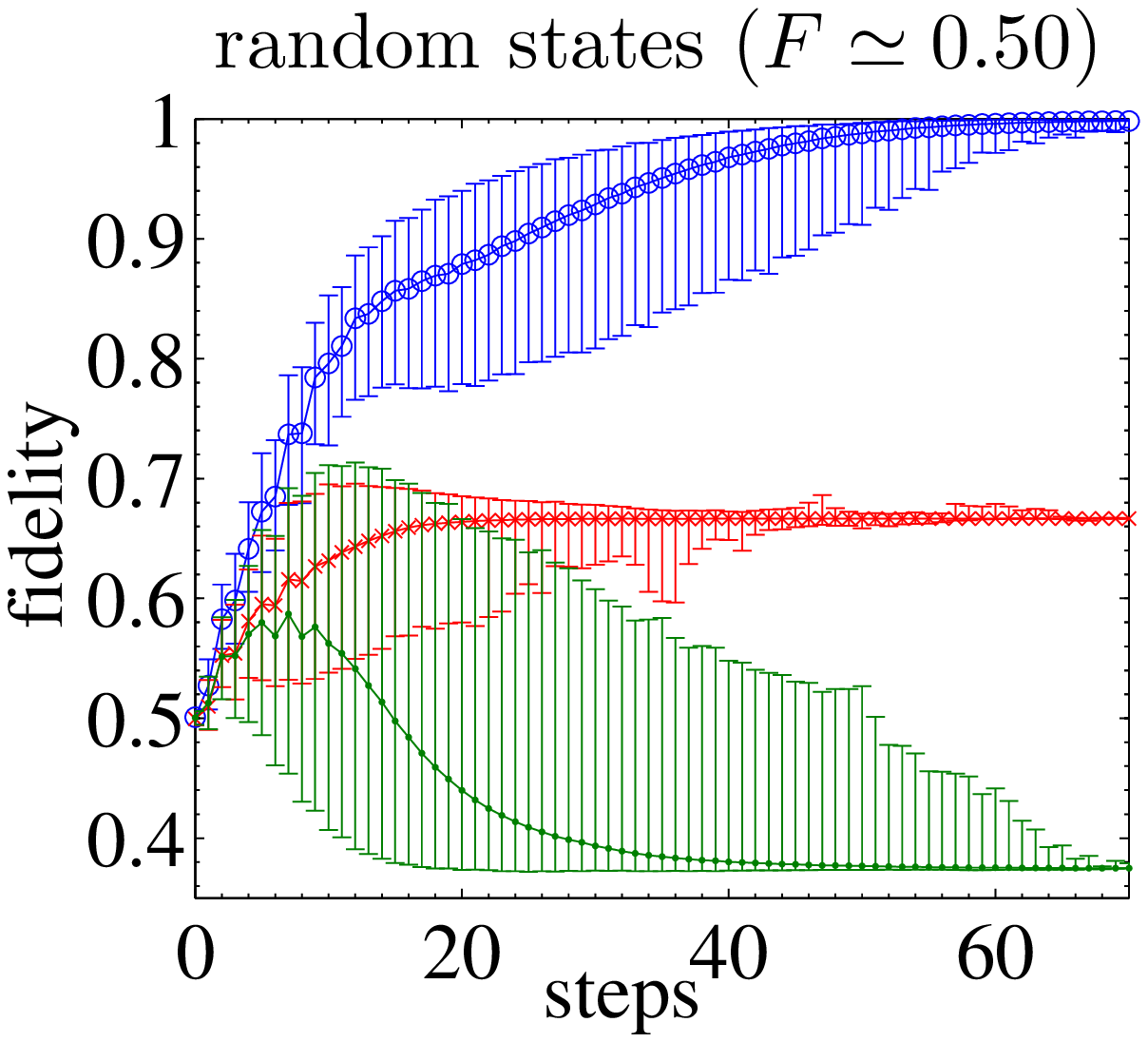}
\end{center}
\end{minipage}
\caption{The average fidelity and its standard deviation followed toward 
each fixed point, for (in total) 10,000 randomly generated initial mixed states 
with $F \in 0.70 \pm 0.01$ (left) or $0.50 \pm 0.01$ (right).}
\label{fig:random}
\end{figure}

Next, we consider randomly generated input mixed states 
(under the Hilbert-Schmidt measure \cite{zyczkowski01}), and 
will observe numerically hierarchical distillations not only to the 3-qubit 
W state, but also to a 2-qubit Bell pair.
This is surprising, since it implies that we can distill
a non-stabilizer state and a stabilizer state by the same protocol.
In Fig.~\ref{fig:random}, for 10,000 random mixed states with the initial 
fidelity $F$ fixed close to 0.70 or 0.50,
we display the average fidelity and its standard deviation for each set of
samples reaching the same fixed point. 
When $F$ is sufficiently large, such as $F\simeq 0.70$, 
the branch to the W state is dominant. 
More than 99 percent of the states follow it, and a few residual samples 
are transient or drop to other fixed points mentioned below.
As $F$ becomes smaller, there appear
three hierarchical branches (i) to the W state, (ii) to the 2-qubit Bell state
($F=\tfrac{2}{3}$) shared by two parties out of three, i.e.,
$\tfrac{1}{\sqrt{2}}(|01\rangle +|10\rangle)|0\rangle^{{l_1}{l_2}{l_3}}$,
where $(l_1,l_2,l_3)$ is a permutation of (A, B, C), or (iii) 
to the undistillable state $\chi$ (up to local unitaries).
Depending on the initial entanglement, a branch is selected by the protocol.
This hierarchy reflects the ``onion-like'' geometry among different kinds of 
entanglement in 3-qubit mixed states \cite{acin01}.
As $F$ approaches the ``critical'' region ($F \simeq 0.50$ in 
Fig.~\ref{fig:random}) for distillability, 
the characteristic number of steps toward every fixed point as well as 
the fluctuation of the progresses for different samples become larger.
The fraction of the states which follow lower branches also increases.

{\it Conclusion.--}
Identifying a complementary (mutually unbiased) pair of stabilizer 
measurements, which replaces the conventional bilateral {\sc cnot}, as a key 
local operation for distillation, 
we have proposed a 3-qubit W state distillation protocol. 
To our knowledge, it is the first protocol to distill directly 
multipartite non-stabilizer states. 
An extension to the $n$-qubit W state should be straightforward, introducing
the general W basis by 
$U^{\rm W basis} = \tfrac{1}{\sqrt{n}}\sum_{l=1}^{n} 
Z^{1}\cdots Z^{l-1} X^{l} \openone^{l+1} \cdots \openone^{n}$.
Since our protocol distills a non-stabilizer state and stabilizer
states on the same footing, our scheme may lead to a unified 
construction of direct distillation protocols for multipartite entanglement.
It is still open whether a hashing protocol can be made
for non-stabilizer states without local depolarization which makes
density matrices classical mixtures of pure states.
Finally, since quantum computers in which only stabilizer states are generated 
can be efficiently simulated by classical computers \cite{gottesman98},
the appearance of non-stabilizer states, such as the W state, is necessary
to exploit the power (universality) of quantum computers. 
Thus, the technique to purify such states, beyond the ``classical'' parity 
check (exclusive {\sc or} via {\sc cnot}) for stabilizer states, might also 
give a new perspective on fault-tolerant quantum computation 
(cf. Ref.~\cite{bravyi05}).

We thank J. Calsamiglia, W. D\"{u}r, O. G\"{u}hne, M. Hein, C. Kruszynska, 
and J. Shimamura for helpful discussions.
This work was supported in part by 
the Japan Society for the Promotion of Science (JSPS),
the Austrian Science Foundation (FWF), and 
the European Union (IST-2001-38877, -39227, SCALA).

\end{document}